# Momentum-space modulated symmetries in the Luttinger liquid

Alexandre Chaduteau ,* Nyan Raess,† Henry Davenport , and Frank Schindler 
*Blackett Laboratory, Imperial College London, London SW7 2AZ, United Kingdom*

The chiral Luttinger liquid develops quantum chaos as soon as a—however slight—nonlinear dispersion is introduced for the microscopic electronic degrees of freedom. For this nonlinear version of the model, we identify an infinite family of translation-invariant interaction potentials with corresponding modulated symmetries. These symmetries are highly unconventional: they are modulated in momentum space (and do not seem to have an easy physical interpretation). We develop a systematic understanding of these symmetries and study the resulting blocks in the Hamiltonian. In particular, this approach allows us to predict the analytic Hamiltonian block sizes and derive asymptotic scaling laws in the limit of large total momentum. These blocks are reminiscent of Hilbert space fragmentation in that, even though they are labeled by a symmetry, this symmetry is highly nonlocal and does not have a simple interpretation. We corroborate this result by studying entanglement entropy and level statistics.

## I. INTRODUCTION

Systems with unconventional symmetries are of current interest because they can display a wide range of interesting properties such as Hilbert space fragmentation [1–3], UV/IR mixing [4,5], subdiffusive transport in nonequilibrium systems [6–11], and the emergence of fractons [12]. Amongst the unconventional symmetries that have been studied, dipole and multipole-moment conserving systems have attracted particular attention because they can exhibit Hilbert space fragmentation. These symmetries as well as subsystem symmetries are examples of the more general class of *spatially modulated* symmetries, which are characterized by conserved quantities of the form $T_{\{\alpha_r\}} = \sum_r \alpha_r q_r$ where $q_r$ is a local operator at position $r$ [13–15].

Inspired by the rich physics found in systems with spatially-modulated symmetries, in this paper we study a celebrated model—the Luttinger liquid with an integrability-breaking perturbation—that can exhibit *momentum-space* modulated symmetries. Analogous to spatially modulated symmetries, these take the form $T_{\{\alpha_p\}} = \sum_p \alpha_p q_p$, where $q_p$ is local in momentum space. Specifically, we find that the chiral nonlinear Luttinger liquid (CNLLL) can exhibit these symmetries [16–20]. The CNLLL is a 1D model of interacting fermions with a nonlinear but unidirectional dispersion and we find the unconventional symmetries for certain choices of fermion-fermion interaction potential. These symmetries exist in the nonintegrable regime of the CNLLL, when the dispersion relation is nonlinear but still unidirectional, away from the two well-known integrable limits: (1) a free fermion limit where the fermion-fermion interaction is removed and (2) a free boson limit where the dispersion relation is linear [16]. A recent study [21] uncovered families of exact Hamiltonian eigenstates in this system, again for certain choices of the fermion-fermion interaction potential. These exact states share a core property of QMBS [22], in that they have characteristically lower entanglement entropy compared to other eigenstates, when subdividing the system in *momentum space*.

For choices of fermion-fermion interaction potential, which lead to momentum-space modulated symmetries, we study the resulting dynamically disconnected Krylov subspaces by computing each subspace's momentum-space entanglement entropies and level statistics. Our results are reminiscent of Hilbert space fragmentation. As is typical of fragmented systems, each Hilbert subspace presents its own "band" of entropies in the entanglement spectrum. We obtain a simple positive integer parameter $n$ that labels these modulated symmetries and the different corresponding realizations of the CNLLL. As $n$ increases (with other parameters fixed), the corresponding system has an increasing number of symmetry-resolved blocks, until each eigenstate of the Hamiltonian forms its own disconnected subspace. We find explicit relations for the block sizes and show that, at least in some cases, the block pattern can be predicted analytically.

We note that there are varying definitions of Hilbert space fragmentation used in the literature; in most definitions, the number of Krylov subspaces (hereby referred to just as "blocks") should scale exponentially rather than polynomially in the system size $L$ [23]. In our case, this definition is difficult to apply: The CNLLL is a continuum system, and a finite Hilbert space is not obtained by fixing the system size $L$, but instead by fixing the total momentum $p_{\text{tot}}$. In this paper, we find that the number of blocks scales polynomially in $p_{\text{tot}}$, and the fraction of Hilbert space occupied by the largest block approaches zero in the limit $p_{\text{tot}} \to \infty$. (We also note that some examples of Hilbert space fragmentation with polynomial scaling in $L$ are known [24]). To avoid this ambiguity in definitions, and even though the momentum-space modulated symmetries discussed here are highly nonlocal, we do not interpret our results in terms of Hilbert space fragmentation.

*Contact author: alexandre.chaduteau20@imperial.ac.uk
†Contact author: pnac064@live.rhul.ac.uk

The paper is organized as follows. The necessary background on the CNLLL is given in Sec. II. In Sec. III, we introduce the family of interaction potentials and unconventional symmetry operators whose eigenvalues resolve Hamiltonian blocks. In Sec. IV, we study the thermal structure of resulting Hamiltonians such as the entanglement entropy and the energy level statistics. Lengthy proofs and more detailed numerics are relegated to the appendixes.

## II. MODEL

The Luttinger liquid is a 1D model of interacting fermions, defined here on a circle of circumference $L$. We take the operator $c_x$ to annihilate a spinless electron at position $x$ where $x \in [0, L)$. Then

$$\{c_x, c_y^\dagger\} = \delta(x-y), \quad \{c_x, c_y\} = \{c_x^\dagger, c_y^\dagger\} = 0,$$
$$\{c_p, c_q^\dagger\} = \delta_{pq}, \qquad \{c_p, c_q\} = \{c_p^\dagger, c_q^\dagger\} = 0, \tag{1}$$

where the momentum space ($p$) and real space ($x$) operators are related by Fourier transforms,

$$c_p^\dagger = \frac{1}{\sqrt{L}} \int_{-L/2}^{L/2} \mathrm{d}x\, e^{\mathrm{i}px} c_x^\dagger, \quad c_x^\dagger = \frac{1}{\sqrt{L}} \sum_p e^{-\mathrm{i}px} c_p^\dagger, \tag{2}$$

and $p \in \frac{2\pi}{L}\mathbb{Z}$ due to periodic boundary conditions. We set $L = 2\pi$ from now on so that all momenta are integers. The Luttinger liquid Hamiltonian splits into a kinetic ($H_{\text{kin}}$) and fermion-fermion interaction part ($H_{\text{int}}$),

$$H = H_{\text{kin}} + H_{\text{int}}, \tag{3}$$

where

$$H_{\text{kin}} = \sum_p \epsilon(p) c_p^\dagger c_p,$$
$$H_{\text{int}} = \int_{-L/2}^{L/2} \mathrm{d}x \int_{-L/2}^{L/2} \mathrm{d}y\, V(x-y)\, c_x^\dagger c_x c_y^\dagger c_y. \tag{4}$$

$H_{\text{int}}$ encodes a two-body interaction with a translationally invariant potential $V(x, y) = V(x-y)$, which could be, e.g., of the Coulomb form. We only consider inversion-symmetric potentials i.e., $V(x) = V(-x)$. To guarantee chirality, we assume that the dispersion relation $\epsilon(p)$ obeys the condition $\text{sgn}[\epsilon(p)] = \text{sgn}(p)$ for all momenta $|p| \ll \Lambda$, where $\Lambda$ is a momentum cut-off scale, usually given by a microscopic lattice spacing $d$ as $\Lambda \sim 1/d$. For instance, Eq. (3) could be the effective low-energy Hamiltonian governing the 1D chiral edge mode of a 2D Chern insulator with Chern number $C = 1$, microscopically defined on a lattice with spacing $d$.

In momentum space, the many-body ground state $|\Omega\rangle = \prod_{p\leqslant 0} c_p^\dagger |0\rangle$ of $H_{\text{kin}}$ has all momenta with nonpositive $\epsilon(p)$ occupied ($|0\rangle$ is the absolute fermionic vacuum). To make the theory well-defined, we normal-order (::) all operators $\mathcal{O}$ with respect to this $|\Omega\rangle$ [25], which moves all $c_{p>0}$ and $c_{p\leqslant 0}^\dagger$ to the right of all other operators in $\mathcal{O}$, keeping track of the anticommutation relations of Eq. (1). This results in a shift $:\mathcal{O}: = \mathcal{O} - \langle\Omega|\mathcal{O}|\Omega\rangle$. In particular, normal-ordering shifts the kinetic energy of $|\Omega\rangle$ to zero and allows us to take the limit $\Lambda \to \infty$ without divergences.

The Luttinger Hamiltonian in Eq. (3) has two important conventional symmetries: (1) $U(1)$ phase rotation symmetry, with an associated conserved total particle number $N = \sum_p c_p^\dagger c_p$, and (2) spatial translation symmetry, with conserved total momentum $P = \sum_p p c_p^\dagger c_p$. When normal-ordering, $:N: |\Omega\rangle = 0$ and $:P: |\Omega\rangle = 0$. From now on, we work within the charge-neutral Hilbert space sector of $\langle :N: \rangle = 0$. Also, since total momentum is conserved, we can work within a sector of fixed total momentum $\langle :P: \rangle = p_{\text{tot}}$.

The full Hilbert space is spanned by basis states of the form

$$|\mathbf{n}, \bar{\mathbf{n}}\rangle = \prod_{p>0} c_p^{\dagger n_p} \prod_{p\leqslant 0} c_p^{\bar{n}_p} \, |\Omega\rangle, \tag{5}$$

where $\mathbf{n}$ and $\bar{\mathbf{n}}$ are particle/hole occupation vectors, with entries equal to 0 or 1. Since $:N: |\Omega\rangle = 0$, both $\mathbf{n}$ and $\bar{\mathbf{n}}$ have the same number of nonzero entries. We will use this basis exclusively, and sometimes refer to it as the *fermionic basis*, to distinguish it from the more commonly used bosonic basis for the linear Luttinger liquid [16]. While the bosonic basis diagonalizes both $H_{\text{kin}}$ and $H_{\text{int}}$ of the linear Luttinger liquid where $\epsilon(p) \sim p$, it does not diagonalize the CNLLL where $\epsilon(p) = vp + ap^2 + \cdots$, because in this case $H_{\text{kin}}$ becomes nondiagonal in the bosonic basis. For our purposes, it is easier to work with the fermionic basis, at least as long as we do not restrict to a specific form of the nonlinear dispersion $\epsilon(p)$. Ordering these states by their total momentum eigenvalue, the dimension of the sector with $\langle :N: \rangle = 0$, $\langle :P: \rangle = p_{\text{tot}}$ is $\mathcal{P}(p_{\text{tot}})$, the number of integer partitions of $p_{\text{tot}}$ [16,26].

The normal-ordered interaction Hamiltonian $:H_{\text{int}}:$ can be simplified as [21]

$$:H_{\text{int}}: = \sum_{q>k} \sum_{p>(k-q)/2} [V(q-k+p) - \delta_{p\neq 0} V(|p|)]$$
$$\times c_{q+p}^\dagger c_q c_k c_{k-p}^\dagger, \tag{6}$$

where $V(p) = \frac{1}{L}\int \mathrm{d}x\, e^{ipx} V(x)$ is the Fourier transform of the real-space potential. This form of the Hamiltonian will turn out to be most useful form to investigate the block structure.

In the noninteracting limit $V(x) = 0$, the eigenstates of the full Hamiltonian $:H:$ have a simple form because the kinetic term ($:H_{\text{kin}}:$) is diagonal in momentum space. We label the eigenstates of $:H_{\text{kin}}:$ in each total momentum sector ($p_{\text{tot}}$) as $|\phi_i\rangle$. For example for $p_{\text{tot}} = 3$ we have the three states

$$|\phi_1\rangle = c_3^\dagger c_0 |\Omega\rangle, \quad |\phi_2\rangle = c_2^\dagger c_{-1} |\Omega\rangle, \quad |\phi_3\rangle = c_1^\dagger c_{-2} |\Omega\rangle. \tag{7}$$

To declutter the notation, we will drop all normal ordering symbols (::) from now on, and implicitly assume that all operators are properly normal ordered.

## III. SYMMETRY-RESOLVED PATTERNS

Each realization of the CNLLL requires specifying the nonlinear dispersion $\epsilon(p)$ as well as the Fourier components of the interaction potential $V(p)$. We define a family of potential

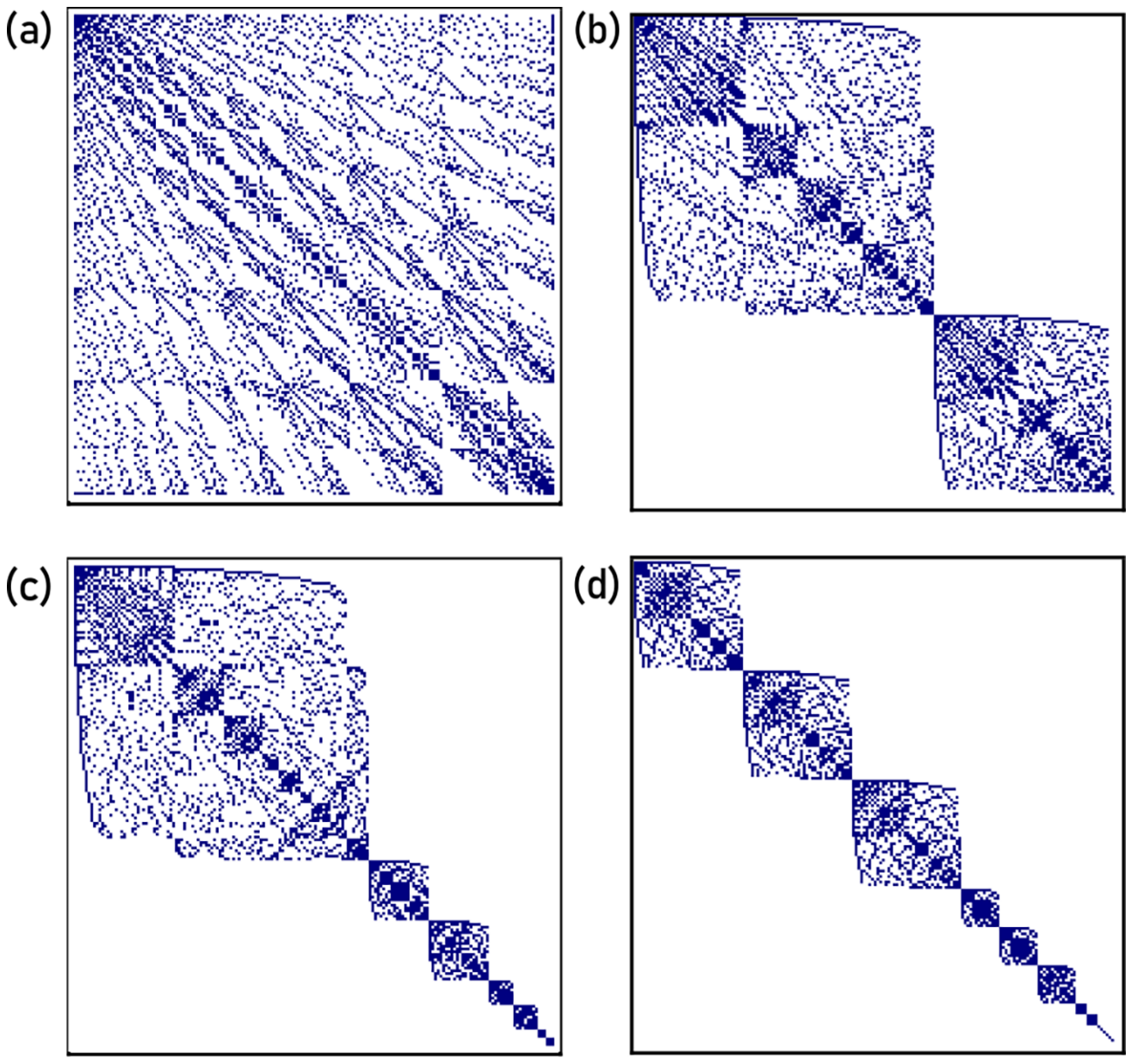

FIG. 1. Fermionic basis matrix representations of $:H_{\text{int}}:$ for $p_{\text{tot}} = 15$; $v = 1$, $a = 0.1/p_{\text{tot}}$ in $\epsilon(p) = vp + ap^2$. Block patterns corresponding to different unconventional symmetries $T_n$ [Eq. (10)] are shown; white and navy pixels represent zero and nonzero matrix elements respectively. The Hilbert space dimension is given by the number of partitions $\mathcal{P}(15) = 176$. (a) $T_1$: No constraints are enforced on the interaction potentials $V_p$ and so the matrix does not block diagonalize. (b) $T_2$: every potential $V_p$ with $p$ odd is set equal and the matrix splits in two blocks. (c) $T_3$: every potential with $p$ not a multiple of 3 is set equal. (d) $T_4$: every potential with $p$ not a multiple of 4 is set equal. In all cases, $V(p) = 0.1$ for all Fourier modes included in the pattern $P_n$, otherwise they are randomly drawn from the interval [0.05, 0.15].

patterns $V(p)$, where we set equal all elements in the set

$$P_n = \{V(p)\,|\, p \in [1, p_{\text{tot}}],\ p \in \mathbb{Z} - n\mathbb{Z}\}. \tag{8}$$

For any choice of chiral dispersion $\epsilon(p)$, these choices of $V(p)$ correspond to a modulated symmetry $T$. For example, $P_2 = \{V(1), V(3), ...\}$ corresponds to setting all odd-momentum Fourier components equal.

An example of the block-diagonal structure of the interaction Hamiltonian matrix elements $\langle\phi_i|H_{\text{int}}|\phi_j\rangle$ for $V(p)$ observing $P_2$ and $P_3$ for $p_{\text{tot}} = 15$ is shown in Fig. 1. Note that there are more blocks for $P_3$ than $P_2$; this is because $P_3$ implies a larger number of constraints. In general, as $n$ increases, more and more blocks are produced at a given $p_{\text{tot}}$. This proceeds until the Hamiltonian becomes fully diagonal at $n = p_{\text{tot}} + 1$. The resulting blocks can be labeled by the eigenvalues of a family of unconventional symmetry operators $T$ that commute with the Hamiltonian.

### A. Modulated symmetries

For each potential tuning pattern $P_n$, there exist two (Hermitian) symmetry generators that commute with the full CNLLL Hamiltonian in Eq. (3), defined by

$$T_n^{(1)} = \sum_p \cos\left(\frac{2\pi}{n}p\right)c_p^\dagger c_p, \quad T_n^{(2)} = \sum_p \sin\left(\frac{2\pi}{n}p\right)c_p^\dagger c_p. \tag{9}$$

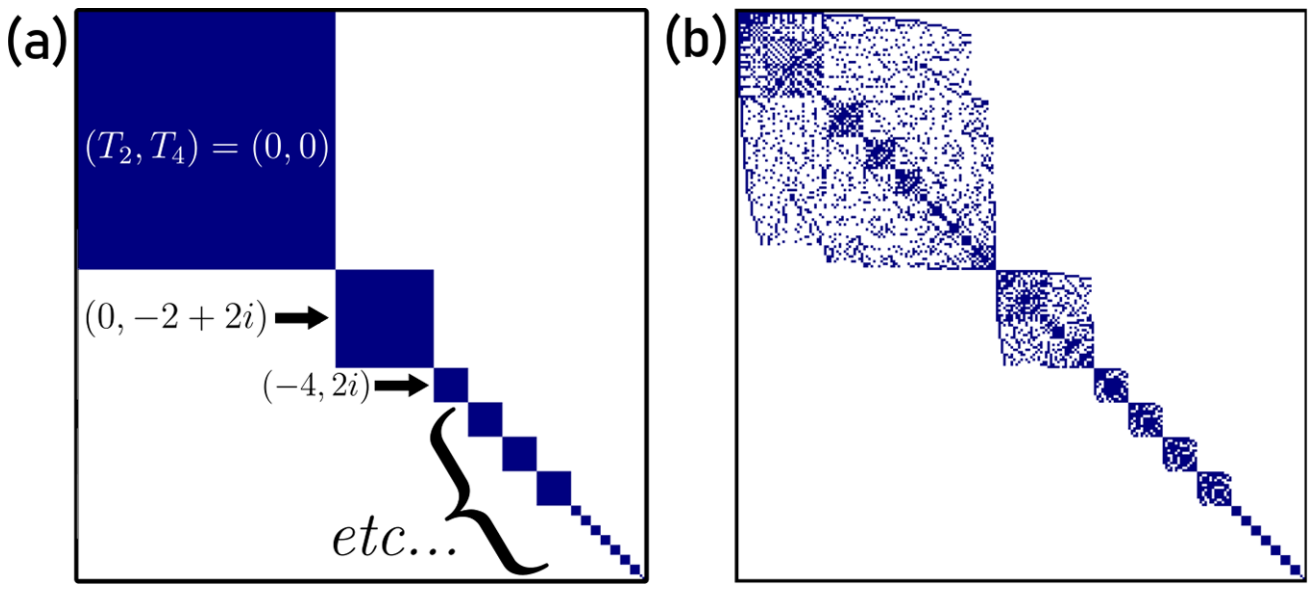

FIG. 2. (a) Block pattern of $H$ for $p_{\text{tot}} = 16$ when grouping basis states by their set of eigenvalues of operators $(T_2, T_4)$. (b) Hamiltonian for $p_{\text{tot}} = 16$ and the $P_4$ pattern, after block diagonalization. Numerical parameter values are the same as in Fig. 1. We set $V(p) = 0.1$ for modes in the $P_4$ pattern; all other modes are drawn randomly from [0.05, 0.15]. The blocks in (a) and (b) match.

We explicitly prove this statement in Appendix A. For example, $T_1^{(1)} = N$ is the number operator and $T_1^{(2)} = 0$. The first nontrivial operator is $T_2^{(1)} = \sum_p (-1)^p c_p^\dagger c_p$, which counts $n_{\text{e}} - n_{\text{o}}$, the occupation number of even momenta states minus odd momenta states, while $T_2^{(2)} = 0$ still only generates the identity. All fermionic states of the form (5) are eigenstates of these symmetry operators. It is practical to package these two operators into a single operator

$$T_n = \sum_p e^{\text{i}\frac{2\pi}{n}p} c_p^\dagger c_p, \tag{10}$$

that is neither Hermitian nor unitary. The Hamiltonian blocks are then labeled by the eigenvalues of $T_n$. In fact, given a set of $V(p)$'s observing $P_n$, not only does $T_n$ commute with $H$, but also $T_f$ for any integer factor $f$ of $n$. If $\{f_1, f_2...\}$ is the set of factors of $n$, then each block is uniquely identified by a set of eigenvalues $\{\lambda_{f_1}, \lambda_{f_2}...\}$, one for each operator $T_{f_i}$ (see Fig. 2 for an example).

The eigenvalues of the $T_n$ operator are always some linear combination of the $n$-th roots of unity; an example is given in Fig. 3. At $p_{\text{tot}} = 3$, the state $|\phi_2\rangle = c_2^\dagger c_{-1}|\Omega\rangle$ from (7) has $T_2$ eigenvalue 2, while $|\phi_1\rangle$ and $|\phi_3\rangle$ have eigenvalue $-2$. Correspondingly, when all $V(p)$ in $P_2$ are set equal [so that $V(1) = V(3) \neq V(2)$], only $|\phi_1\rangle$ and $|\phi_3\rangle$ interact with one another, while $|\phi_2\rangle$ remains as an isolated "scar" state. As for $T_3$, since all three states have different $T_3$ eigenvalues, the matrix becomes diagonal when $P_3$ is enforced; all states are disconnected.

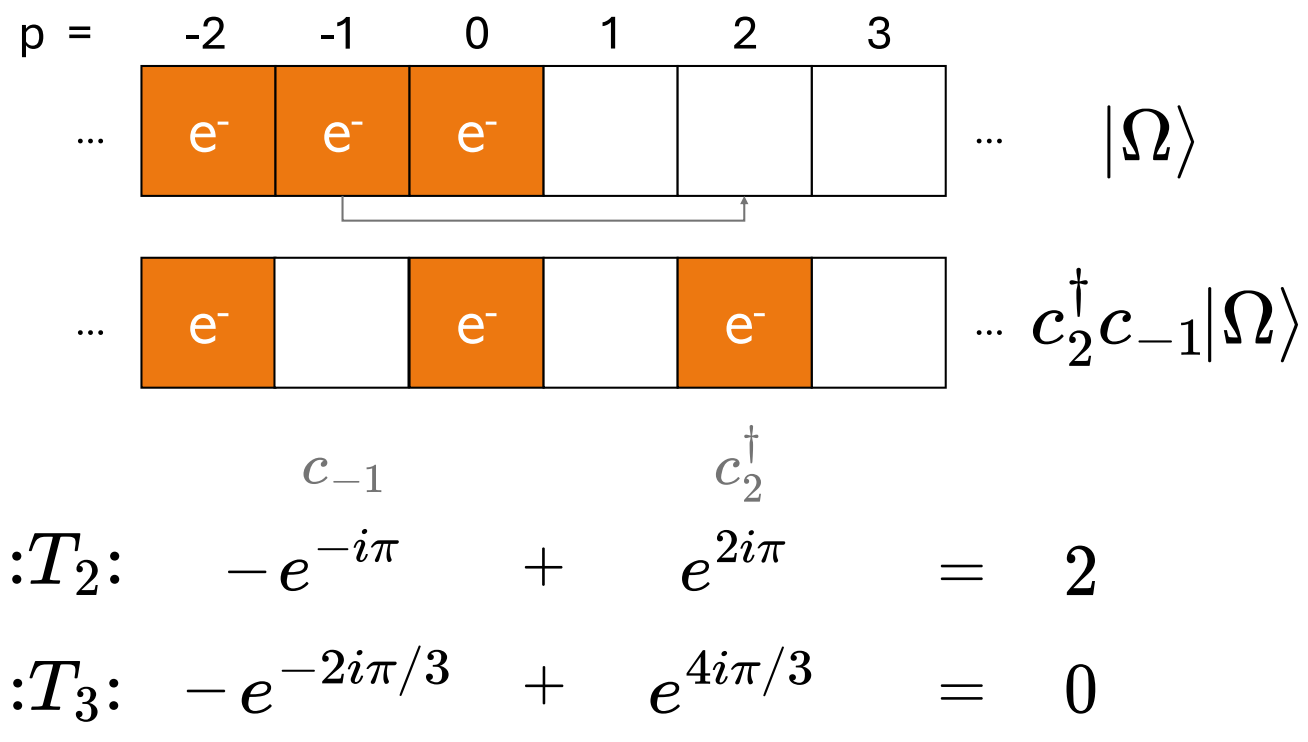

FIG. 3. Example calculation of $:T_2:$ and $:T_3:$ eigenvalues of the state $|\phi_2\rangle = c_2^\dagger c_{-1}|\Omega\rangle$ at $p_{\text{tot}} = 3$. Fermionic commutation relations determine the sign of each term; each of the operators $c_2^\dagger$ and $c_{-1}$ contributes as shown.

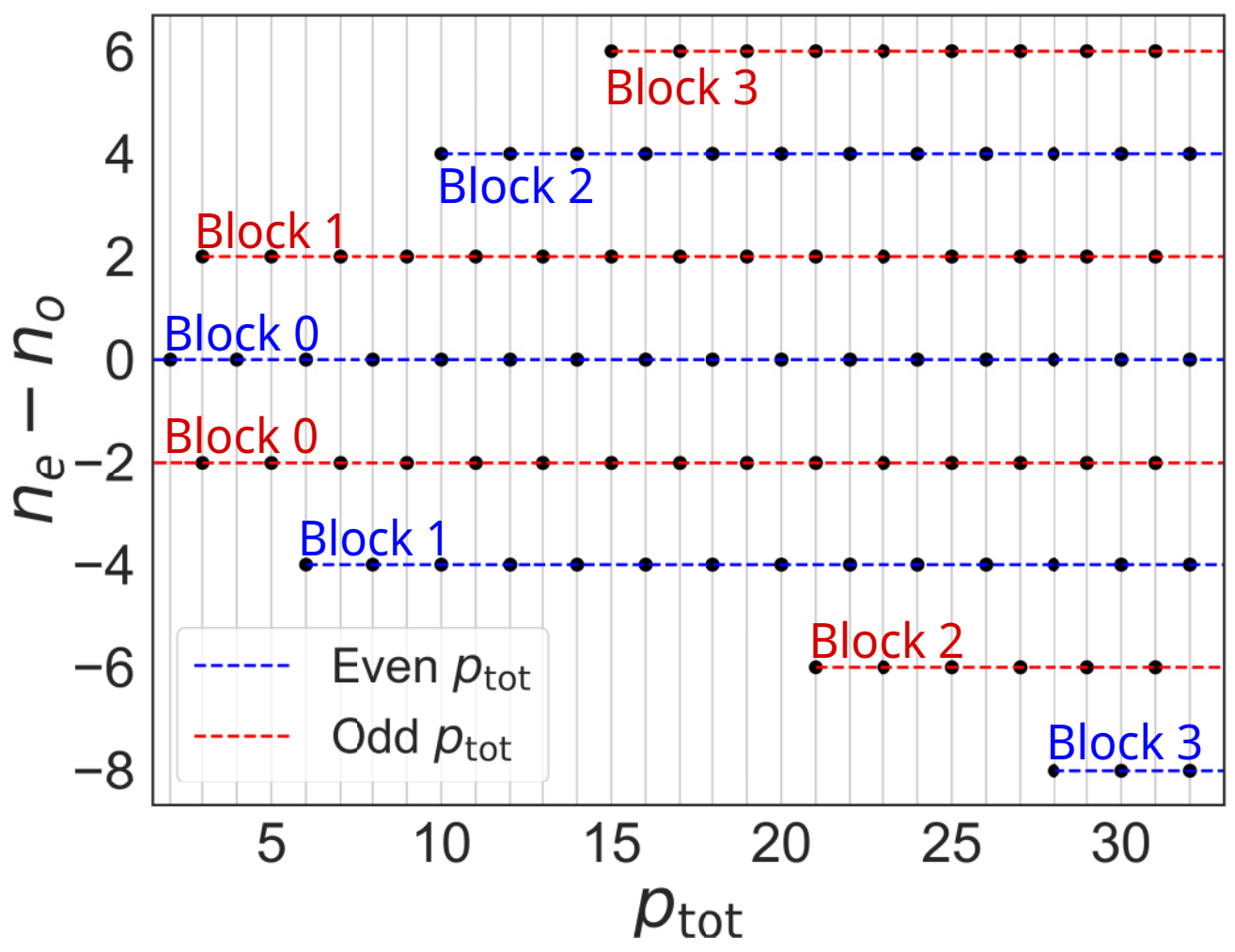

FIG. 4. $T_2$-resolved Hamiltonian blocks and their eigenvalues $n_e - n_o$ against $p_{tot}$. New blocks arise exactly at triangular numbers $t_i$ [Eq. (12)]. Note that there are families of blocks with the same eigenvalues for a range of $p_{tot}$ even (blue) and $p_{tot}$ odd (red).

The number of different eigenvalues of $T_n$ that appear is dependent on the $p_{tot}$ subsector (see Fig. 4). For example, at $p_{tot} = 6$, the state $c_{-2}c_0c_1^\dagger c_3^\dagger|\Omega\rangle$ has $T_2$ eigenvalue $-4$, but at smaller total momenta there is no state with this eigenvalue. We say that this eigenvalue is "new" at $p_{tot} = 6$. More precisely, there is a "new" block at total momentum $p$ if the set of eigenvalues $\{\lambda_{f_1}, \lambda_{f_2}...\}$ that label it is not shared by any state in subsectors $p_{tot} < p$. Conversely, the remaining blocks can be uniquely identified with blocks appearing for smaller $p_{tot}$, as they share the same eigenvalues $\{\lambda_{f_1}, \lambda_{f_2}...\}$.

### B. Analytical results for block sizes

For the sequence of symmetries $T_n$ with $n$ prime, our numerical results suggest that blocks "grow" according to the sequence $a_n(j)$, the number of partitions of $j$ into integers of $n$ kinds (for $n = 2$, see Ref. [27]). For instance, $a_2(2) = 5$ because, labeling the two kinds of integer as primed and unprimed, we have

$$2 = 2' = 1 + 1 = 1' + 1' = 1 + 1'$$

giving five possible partitions. If a block is "new" at $p_{tot} = p$, it will be of size $a_n(j)$ at $p_{tot} = p + nj$ (where $n$ is fixed by the assumption of symmetry $T_n$ and the statement holds for any choice of non-negative $j$). Since $a_n(0) \equiv 1$, a new block always consists of a single state; equivalently, at each $p_{tot}$ at which a new eigenvalue occurs, there is only one state with that eigenvalue. As a minimal example, take the symmetry $T_2$; the block with $T_2$ eigenvalue 0 begins at $p_{tot} = 0$ with only the ground state $|\Omega\rangle$, and at $p_{tot} = 4$ this block contains 5 states. The generating function of the sequence $a_n(j)$ is

$$z_n(x) = \prod_{k\geqslant 1} \frac{1}{(1 - x^k)^n}, \tag{11}$$

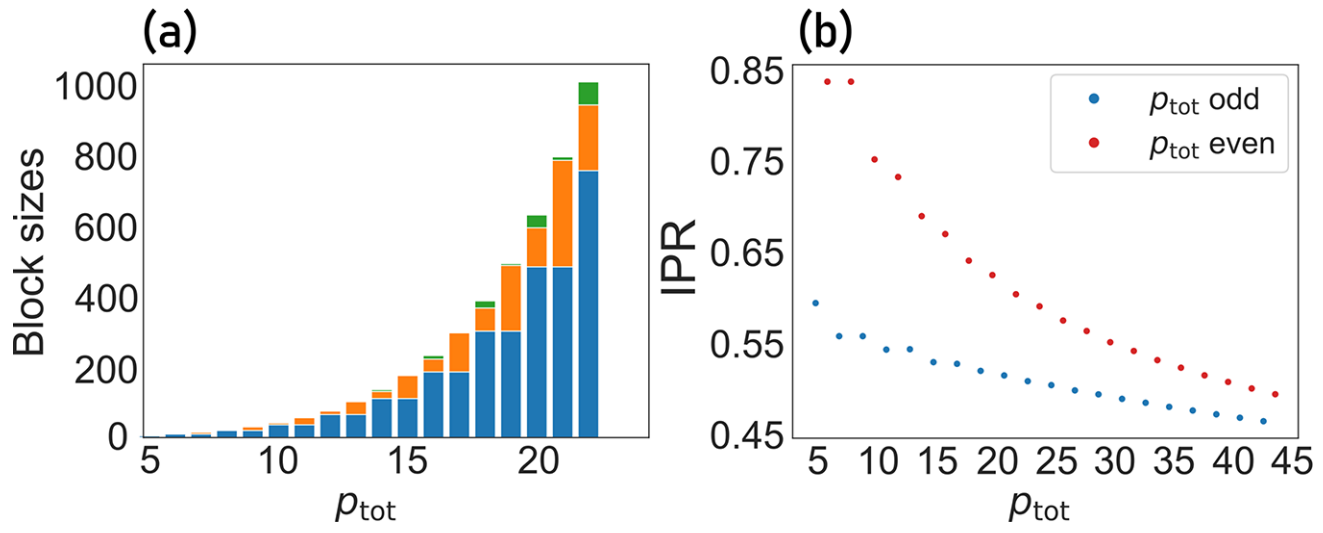

FIG. 5. (a) Stacked-bar histograms of block sizes of the Hamiltonian matrix at a given $p_{tot}$, for symmetry $T_2$. Each color represents a different block; the height of a whole bar is the full Hilbert space dimension. (b) Inverse participation ratio [IPR, Eq. (13)] of the block distribution against $p_{tot}$ for $P_2$. A high IPR implies a peaked distribution (only a few blocks dominate the full Hilbert space), while a low IPR implies a broad distribution (many blocks coexist with roughly similar sizes). We color-code the $p_{tot}$ that are odd or even.

meaning that $z_n(x) = a_n(0) + a_n(1)x + a_n(2)x^2 + a_n(3)x^3 + \cdots$. For the symmetry $T_2$ we give a proof that $a_2(j)$ calculates the block sizes in Appendix E. For $n \neq 2$ the above remains a conjecture to which we have not found a numerical counterexample for $n \leqslant 10$ and $p_{tot} \leqslant 30$. Numerically, we have observed that the series $a_n(j)$ also predicts block sizes for $n$ nonprime, i.e., even when further symmetries $T_f$ for all factors $f$ of $n$ are present.

Figure 5 shows the block populations for $T_2$, whose block sizes are predicted by sequence $a_2$. The number of blocks scales polynomially with $p_{tot}$. This should be contrasted with a definition of Hilbert space fragmentation where the number of blocks scales exponentially with the system size $L$ [28], see our discussion in Sec. I. We include a similar plot for $T_3$ in the Appendix.

For the symmetry $T_2$ we can also predict the values of $p_{tot}$ at which new Hamiltonian blocks arise: the $T_2$ operator gains a new eigenvalue (i.e., there is a new block) whenever $p_{tot}$ reaches a triangular number, where the triangular numbers, $t_i$, are defined as the sum of the first $i$ nonzero integers,

$$t_i = \sum_{k=1}^{i} k = \frac{i(i+1)}{2}. \tag{12}$$

We call the special state that has the new eigenvalue at $p_{tot} = t_i$ a *triangular state* $|\Delta_i\rangle$, with eigenvalue $\Delta_i$. Figure 4 shows new blocks arising at triangular numbers, as well as their associated eigenvalues. Any eigenvalue that is new has the largest possible magnitude in its $p_{tot}$ subspace; we motivate this in Fig. 6. As $n$ increases new blocks quickly start to arise in almost every $p_{tot}$ sector.

Numerically, for $n = 3$ blocks arise at numbers matching OEIS sequence A267137 [29] (apart from the second entry of the sequence, which could be considered an edge/convention issue). This observation could help in proving the conjecture for higher $n$.

Finally, we note that our block size conjecture relates blocks at $p_{tot}$ with those at $p_{tot} + nj$; numerically, we know this is because, if $T_n$ has some set of eigenvalues $\{\lambda_{n_1}, \lambda_{n_2}, \lambda_{n_3}...\}$ labeling blocks in a $p_{tot}$ subsector, those eigenvalues will next appear in the $p_{\mathrm{tot}} + n$ subsector (see Fig. 4 for $T_2$ where eigenvalues reappear whenever $p_{\mathrm{tot}}$ is advanced by 2). This means we can consider each of the $n$ '$p_{\mathrm{tot}}$ mod $n$' sequences separately (Appendix E).

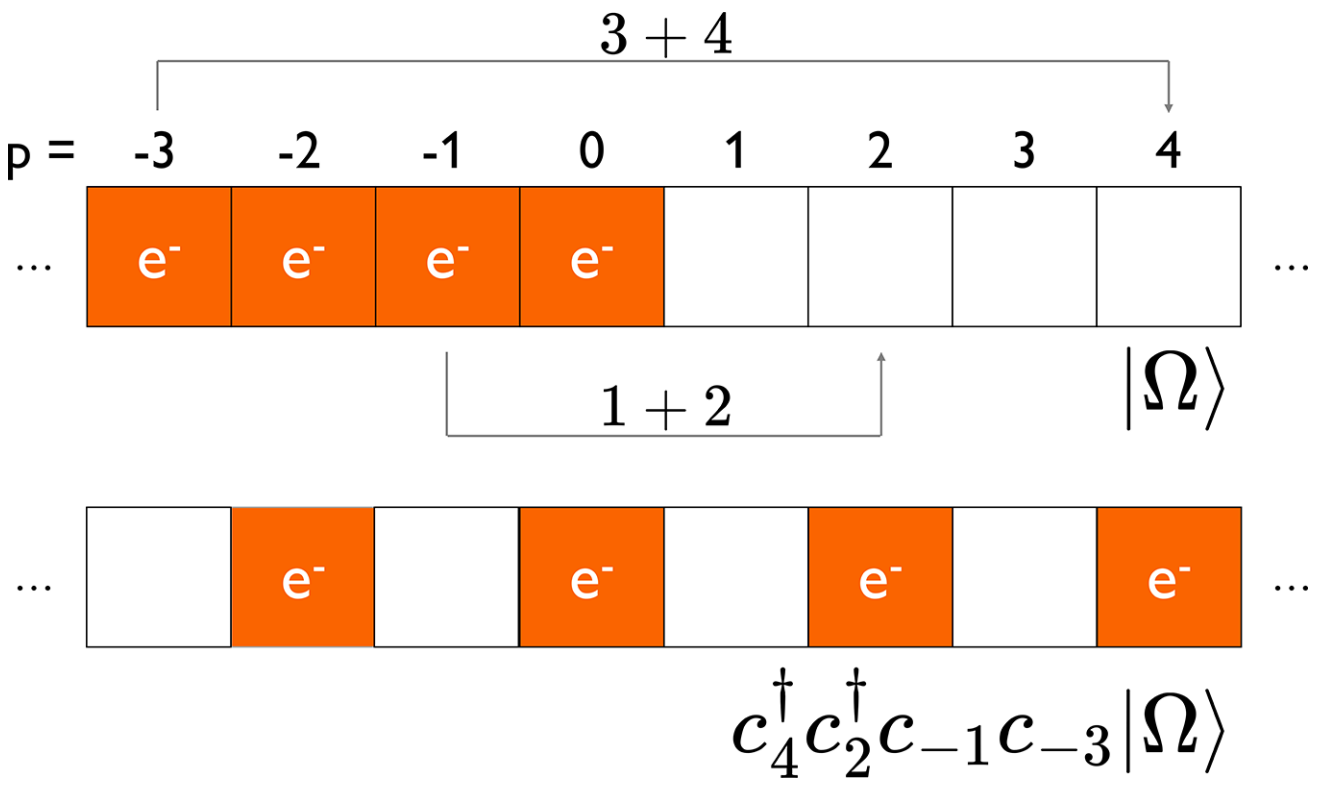

FIG. 6. A triangular state at $p_{\mathrm{tot}} = 10$. Any eigenvalue of the $T_2$ operator may be expressed as $\lambda = n_{\mathrm{e}} - n_{\mathrm{o}}$, where $n_{\mathrm{e}}$ and $n_{\mathrm{o}}$ denote the number of even and odd momentum sites occupied. To maximise $|\lambda|$ while minimizing $p_{\mathrm{tot}}$ one must remove the first available odd/even electron sites and replace them at even/odd sites. As shown above for the case $10 = 4 + 3 + 2 + 1$, this always results in a triangular number.

### C. Numerically observed block distribution

We here further study the number of blocks and their relative size against $p_{\mathrm{tot}}$. To do so, we measure the inverse participation ratio (IPR), defined as

$$\mathrm{IPR} = \sum_i \frac{d_i^2}{D^2} \tag{13}$$

where $d_i$ is the dimension of the $i$th block and $D$ is the size of the full Hilbert space at a given $p_{\mathrm{tot}}$. There are two characteristic régimes: (1) there is a dominant block and the IPR is almost 1; (2) there is no dominant block, there are many blocks, and the IPR approaches 0. To seek the "most fragmented" pattern then corresponds to minimizing the IPR. At a given $p_{\mathrm{tot}}$, the potential configuration that minimises the IPR is the one with the most potentials set equal [Eq. (8)], simply because this destroys the most off-diagonal elements in the corresponding Hamiltonian matrix.

We compute the IPR for a given symmetry $T_n$ against $p_{\mathrm{tot}}$. The result is shown in Fig. 5(b) for $n = 2$. We find that, fixing a symmetry $T_n$, the IPR decreases as $p_{\mathrm{tot}}$ increases, indicating that there is no dominant block. Moreover, the IPR plots for symmetry $T_n$ decompose into $n$ monotonously decreasing curves corresponding to different values of $p = p_{\mathrm{tot}} \pmod n$, with higher $p$ corresponding to lower IPR. This behavior can be qualitatively understood by computing the fraction $f$ of potentials $V(p)$ set equal [Eq. (8)] to enforce a given symmetry $T_n$. $f$ is a measure of how constrained the dynamics are, and should therefore correlate with the number of blocks. For $p_{\mathrm{tot}} = kn + p,\ k \in \mathbb{Z},\ p \in [1, n-1]$, we have $f = \frac{k(n-1)+p}{kn}$, which increases with $p$.

Additionally, we can use the sequence $a_2(j)$ to obtain an upper bound for the scaling of the IPR for $n = 2$. Using the asymptotic scaling given on the sequence's OEIS page [27],

$$a_2(j) \sim \frac{e^{2\pi\sqrt{j/3}}}{4 \cdot 3^{3/4} \cdot j^{5/4}}\left[1 + \mathcal{O}\left(\frac{1}{j^{1/2}}\right)\right], \tag{14}$$

it follows that the IPR for $n = 2$ is bounded from above,

$$\mathrm{IPR}_2(p_{\mathrm{tot}}) \leqslant \frac{2^{5/4}3^{1/2}}{p_{\mathrm{tot}}^{1/2}} + \mathcal{O}\left(\frac{1}{p_{\mathrm{tot}}^{7/4}}\right), \tag{15}$$

and tends to zero as $\sim 1/\sqrt{p_{\mathrm{tot}}}$ when $p_{\mathrm{tot}} \to \infty$. Thus, there is no dominant block in the thermodynamic limit in which $L$ is sent to infinity but the physical momentum (and kinetic energy scale) are held fixed.

For any $n$ nonprime, let $\{f_1, f_2, f_3 ...\}$ be its factors sorted in increasing order. The $T_n$ block pattern can be regarded as first resolving only $T_{f_1}$, then the corresponding blocks "split" into smaller blocks when further resolving $T_{f_2}$, and so on. Suppose each $T_{f_1}$ block of size $d_i$ splits into $T_{f_2}$-resolved blocks $d_{j_i}$ such that $\sum_j d_{j_i} = d_i$. Then the $T_{f_2}$-and-$T_{f_1}$-resolved IPR will be $\sum_j \sum_i d_{j_i}^2 / D^2$. But by the triangle inequality, $\sum_j d_{j_i}^2 \leqslant (\sum_j d_{j_i})^2 = d_i^2$, so this IPR will necessarily be smaller than the $T_{f_1}$-only-resolved IPR. Thus, the IPR for $n$ nonprime is necessarily smaller than the IPR of any of its factors. Using the above result for $n = 2$, this implies that the block distribution resulting from $T_n$ with any even $n$ will also have an IPR decaying at least as $1/\sqrt{p_{\mathrm{tot}}}$ when $p_{\mathrm{tot}} \to \infty$.

## IV. THERMALISATION IN SYMMETRY-RESOLVED BLOCKS: THE CASE OF $T_2$

We now analyze the thermal properties of the $T_2$-resolved blocks. Using the method outlined in Ref. [21], we compute the distribution of momentum-space entanglement entropies for eigenstates of a generic realization of the CNLLL Hamiltonian that commutes with $T_2$. Figure 7(b) shows the entanglement entropies for $p_{\mathrm{tot}} = 30$. Here, each block in the Hamiltonian corresponds to a distinct entropy band. The rough magnitude of entropies in each band scales with the size of the corresponding block.

We also study the adjacency gap ratio statistics, which improves upon simple energy level statistics [30]. Since $H$ is purely real, the Wigner-Dyson distribution of interest is the Gaussian orthogonal ensemble (GOE), with an average gap ratio $\bar{r} \approx 0.536$, indicating thermalization. In comparison, the Poisson distribution (displayed by systems with no dominant block in the thermodynamic limit) has average $\bar{r} \approx 0.386$. We show the detailed distributions for the symmetry $T_2$ and $p_{\mathrm{tot}} = 30$ in Appendix G. As shown in the overview Fig. 8 for $p_{\mathrm{tot}} = 30$, each separate block closely follows GOE statistics. We choose to show data for the biggest block to avoid effects due to smallness of blocks.

However, as $n$ increases and $p_{\mathrm{tot}}$ is held fixed, the largest block's relative size decreases, implying that the quality of the GOE fit decreases. This explains why the red dots do not converge well in $\bar{r}$. This is a finite-size effect: Plotting the same symmetry $T_n$'s level statistics for increasing $p_{\text{tot}}$ gives a $\bar{r}$ value that always approaches the GOE value.

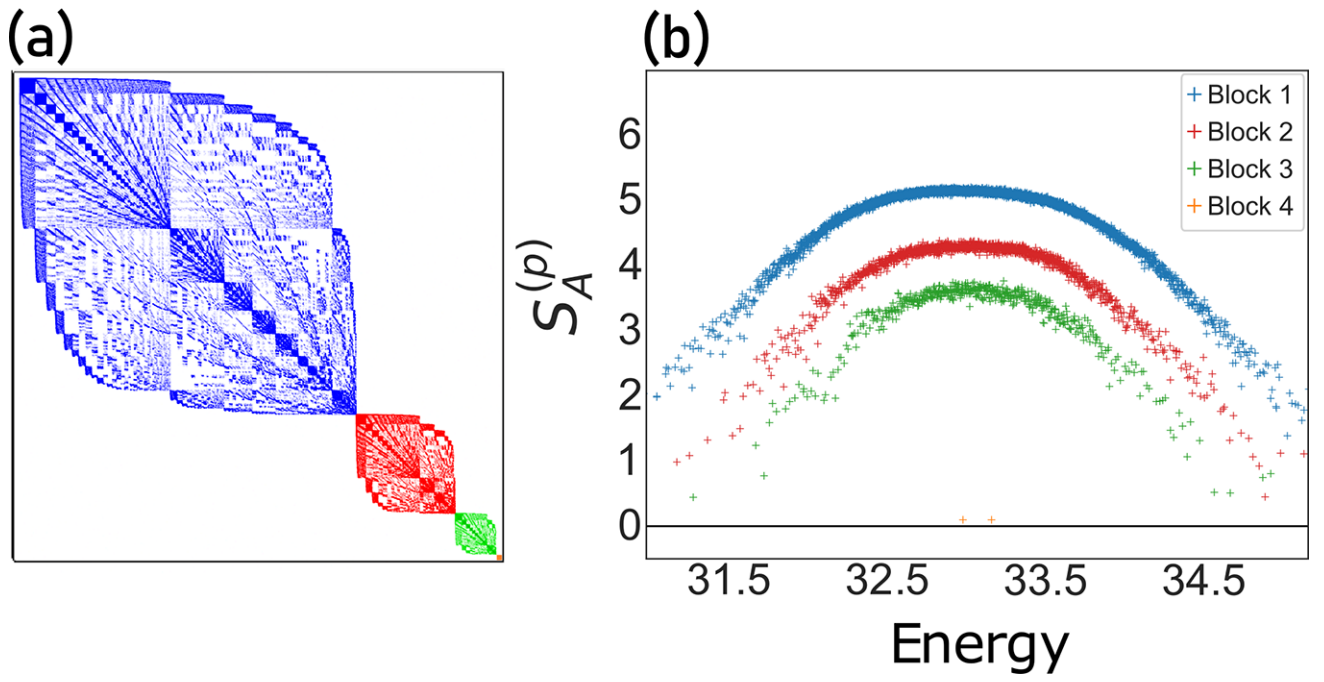

FIG. 7. (a) Hamiltonian matrix for $p_{\text{tot}} = 30$ with symmetry $T_2$ enforced. White cells are zeros; blocks are color coded. (b) Momentum-space entanglement entropies sorted by energy. Parameter values are as in Fig. 1. We divide the single-particle momentum space [the space $p$ ranges over in Eq. (5)] into subspaces $A$ and $B$ that correspond to roughly equal Hilbert space dimensions [21]. Here, for $p_{\text{tot}} = 30$, we choose $A = [-7, 3]$ and $B$ as its complement. These subspaces have Hilbert space dimensions $\dim(\mathcal{H}_A) = 1086$ and $\dim(\mathcal{H}_B) = 996$, while the full Hilbert space has dimension $\mathcal{P}(30) = 5604$ [which is smaller than $\dim(\mathcal{H}_A)\dim(\mathcal{H}_B)$ due to the constraint that the total momentum should be $p_{\text{tot}}$]. Each Hamiltonian block from (a) corresponds to an entropy band in (b); block 4 is too small to be clearly visible.

## V. REAL-SPACE PROFILE

We also consider the real-space profile of the potential configurations corresponding to $P_n$. We show examples for $n = 5$ in Fig. 9 for various choices of values for the strength $V$ of potentials in $P_5$. There are clear nonlocal peaks, although they are smallest when $V$ is at the middle of the interval from which we randomly pick components outside $P_5$. Nonlocality in the potential indicates that the potentials $P_n$ would not correspond to physical vacuum interactions; they could, however, be realized by screening effects in metals.

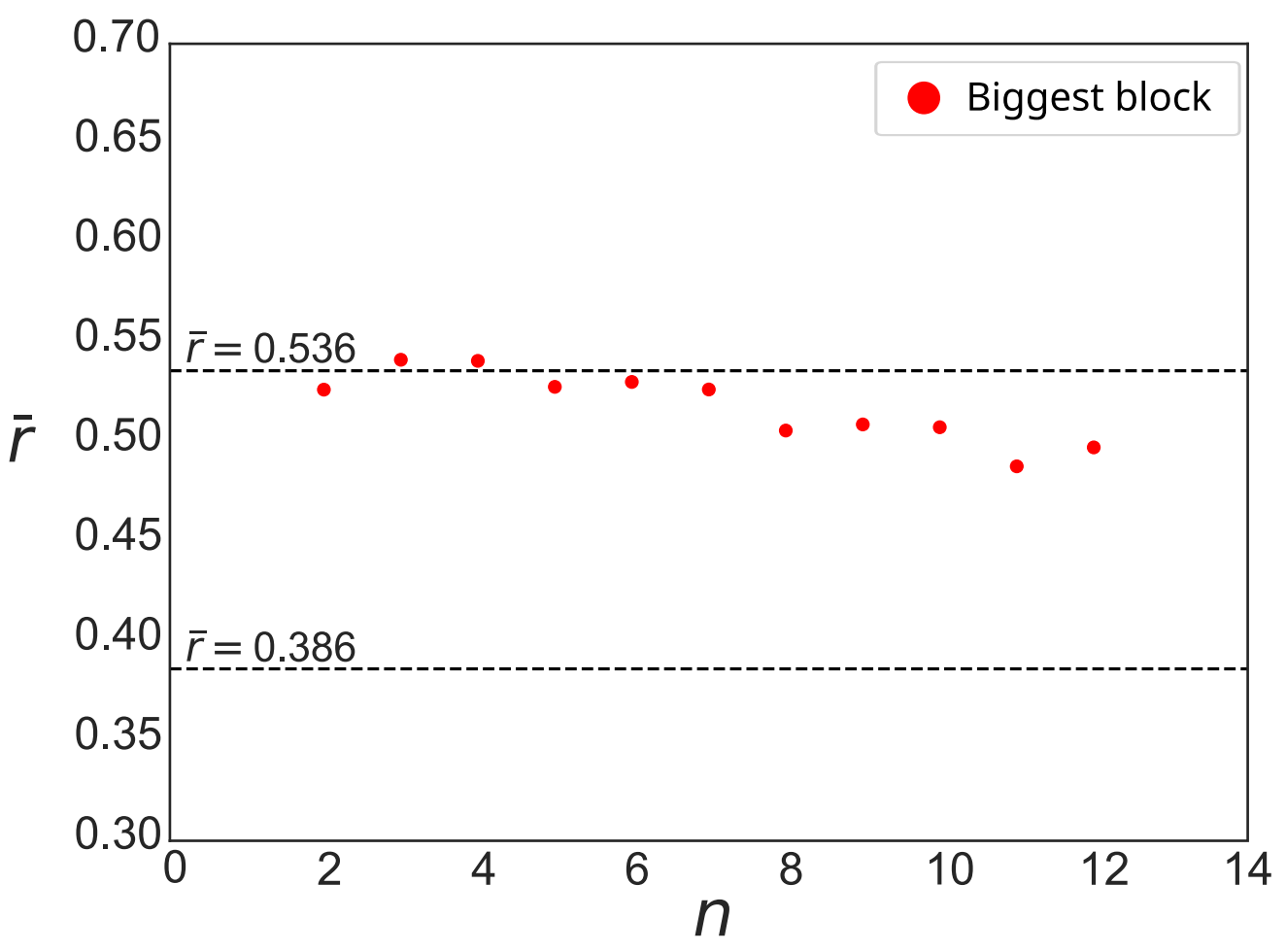

FIG. 8. Average adjacency gap ratio $\bar{r}$ for $p_{\text{tot}} = 30$ when enforcing the various symmetries $T_n$. Numerical parameters are the same as in Fig. 1. Red points represent the biggest Hamiltonian blocks' $\bar{r}$. Dashed-horizontal lines represent the Poisson average $\bar{r} \approx 0.386$ and the (thermal) GOE $\bar{r} \approx 0.536$.

## VI. SUMMARY AND OUTLOOK

We have found an infinite family of unconventional, momentum-space modulated symmetries $T_n$ in the CNLLL. We do not interpret the resulting block structure as Hilbert space fragmentation. However, these symmetries seem to provide a spectrum of various degrees of what, by analogy with Hilbert space fragmentation literature, would be called strong fragmentation between $n = 2$ and $n \to \infty$ (for $p_{\text{tot}} \to \infty$); although the role of finite-size effects in this result is yet to be confirmed. The corresponding Hamiltonians have no dominant symmetry-resolved block in the limit of large total momenta (large Hilbert space sizes). For a fixed instance of the interaction potential $V(p)$, each Hamiltonian block's level statistics are individually of the Wigner-Dyson type, indicating that eigenstates within each block thermalise with each other.

We now list a few open questions. The first is whether there exist other interesting unconventional symmetry patterns in the CNLLL. We performed an exhaustive search (up to $p_{\text{tot}} \leqslant 10$) for simple symmetry patterns beyond those described here, where we set equal all interaction potentials $V(p)$ in the set $P_n$ [Eq. (8)]. Within this range of $p_{\text{tot}}$, the only patterns found were for $V(p)$ potential tuning patterns that are directly inherited from $P_n$ by removing some of the higher-momentum $V(p)$'s.

We must also stress here that we searched for such patterns *only* in the fermionic basis—this amounts to determining whether a given choice of interaction potentials $V(p)$ allows the fermionic-basis Hamiltonian matrix to be permuted into block-diagonal form. The existence of further unconventional symmetry patterns is possible as well and warrants further investigation.

Also, Ref. [23] characterises Hilbert space fragmentation in terms of *commutant algebras*; these are algebras of operators commuting with every term in the Hamiltonian. While as discussed in Sec. I, we do not refer to our results as fragmentation, it could be instructive to explicitly find the commutant algebra for the $V(p)$ potential tuning pattern $P_n$ studied here, which may actually be larger than the algebra of symmetries $T_n$.

## ACKNOWLEDGMENTS

A.C. and N.R. thank Kevin Buzzard for discussions regarding the sequence $a_2(j)$. F.S. thanks Sanjay Moudgalya for help in clarifying the notion of Hilbert space fragmentation. F.S. also thanks Nicolas Regnault and Andrei Bernevig for an earlier collaboration on a similar topic. A.C. acknowledges funding from the Imperial College London President's Ph.D. Scholarships. H.D. acknowledges support from the Engineering and Physical Sciences Research Council (Grant No. EP/W524323/1). F.S. gratefully acknowledges support from the Simons Center for Geometry and Physics, Stony Brook University at which some of the work for this paper was performed. This work was supported by a UKRI Future Leaders Fellowship MR/Y017331/1.

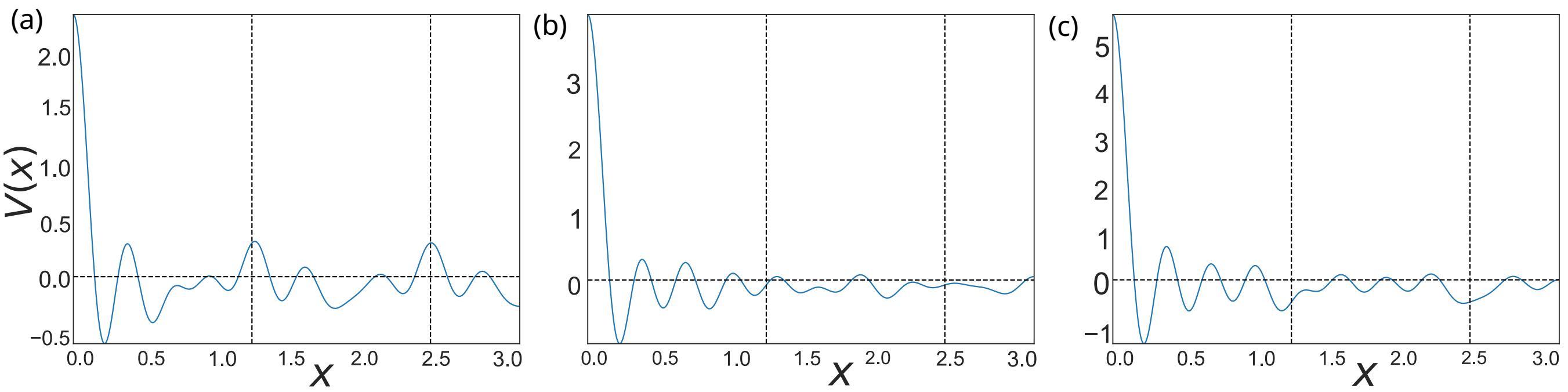

FIG. 9. Real-space potential profiles satisfying symmetry $T_5$, obtained by Fourier-transforming a certain potential configuration $\{V(p)\}$. We restrict to positive fermion separation $x$ since $V(x) = V(-x)$. Potentials not set equal are randomly drawn from the interval [0.05, 0.15]. The potentials set equal take values (a) 0.05, (b) 0.1, (c) 0.15.

## APPENDIX A: MOMENTUM-SPACE COMMUTATOR

Here, we show that the operator $T_n$ becomes a symmetry of the Hamiltonian under pattern $P_n$. Recall that

$$T_n = \sum_p e^{ip\theta_n} c_p^\dagger c_p, \tag{A1}$$

where $\theta_n = 2\pi/n$. The sum goes over all $p \in \mathbb{Z}$, but could be restricted to the range $[-p_{\text{tot}} + 1, p_{\text{tot}}]$ within a given $p_{\text{tot}}$ sector, since roughly speaking, summing all $n$th roots of unity gives zero. To see that $T_n$ is a symmetry, we compute the commutator of $T_n$ with $H_{\text{int}}$ as in Eq. (6) as follows. Consider

$$c_l^\dagger c_l c_{q+p}^\dagger c_q c_k c_{k-p}^\dagger. \tag{A2}$$

We use the anticommutation relations to move $n_l = c_l^\dagger c_l$ to the right end of this string. Doing so and rearranging yields

$$[c_l^\dagger c_l, c_{q+p}^\dagger c_q c_k c_{k-p}^\dagger] = \delta_{l,q+p} c_l^\dagger c_q c_k c_{k-p}^\dagger + \delta_{l,k-p} c_{q+p}^\dagger c_q c_k c_l^\dagger - \delta_{lq} c_{q+p}^\dagger c_l c_k c_{k-p}^\dagger - \delta_{lk} c_{q+p}^\dagger c_q c_l c_{k-p}^\dagger. \tag{A3}$$

Multiplying by $e^{i\theta_n l}$ and summing over the index $l$, we get

$$[T_n, c_{q+p}^\dagger c_q c_k c_{k-p}^\dagger] = c_{q+p}^\dagger c_q c_k c_{k-p}^\dagger \times [e^{i\theta_n(q+p)} + e^{i\theta_n(k-p)} - e^{i\theta_n q} - e^{i\theta_n k}]. \tag{A4}$$

Thus when summing over $q, k$, we have

$$\begin{aligned}[T_n,\ H_{\text{int}}] &= \sum_{q>k} \sum_{p>(k-q)/2} [V(q-k+p) - \delta_{p\neq 0} V(p)] \times [T_n, c_{q+p}^\dagger c_q c_k c_{k-p}^\dagger] \\ &= \sum_{q>k} \sum_{p>(k-q)/2} [V(q-k+p) - \delta_{p\neq 0} V(p)] F_n(q,k,p),\end{aligned} \tag{A5}$$

where we have defined

$$F_n(q,k,p) = e^{i\theta_n(q+p)} + e^{i\theta_n(k-p)} - e^{i\theta_n q} - e^{i\theta_n k}, \tag{A6}$$

and note that $F_n$ vanishes precisely when at least one of $p, q-k+p$ is a multiple of $n$. When $p$ is a multiple of $n$ the first and third term cancel each other, as do the second and fourth; when $q-k+p$ is a multiple of $n$ the first cancels the fourth and the second cancels the third. Hence, for the commutator to vanish completely, the other square bracket term, $V(q-k+p) - \delta_{p\neq 0} V(p)$, must vanish when $p$ and $q-k+p$ are both *not* multiples of $n$. Thus $T_n$ is a symmetry when all $V(p)$ are set equal for all $p$ not multiples of $n$, i.e., when we use pattern $P_n$.

## APPENDIX B: REAL-SPACE INTERPRETATION

Using the Fourier transform of $c_p$ and $c_p^\dagger$ operators, we write

$$\begin{aligned}T_n &= \frac{1}{L} \sum_p \int_{x=-L/2}^{L/2} \mathrm{d}x\, c_x^\dagger e^{ipx} \int_{y=-L/2}^{L/2} \mathrm{d}y\, c_y e^{-ipy} e^{ip\theta_n} \\ &= \frac{1}{L} \int_{x=-L/2}^{L/2} \mathrm{d}x \int_{y=-L/2}^{L/2} \mathrm{d}y\, c_x^\dagger c_y \sum_p e^{ip(x-y+\theta_n)}.\end{aligned} \tag{B1}$$

Using the Dirac comb relation $\sum_{p\in\mathbb{Z}} e^{ipz} = 2\pi \sum_p \delta(z - 2p\pi)$ and $L = 2\pi$, we get

$$\begin{aligned}T_n &= \frac{2\pi}{L} \int_{x=-\pi}^{\pi} \mathrm{d}x \int_{y=-\pi}^{\pi} \mathrm{d}y\, c_x^\dagger c_y \sum_p \delta(x - y + \theta_n - 2p\pi) \\ &= \sum_q \int_{-\pi}^{\pi} \mathrm{d}x\, c_x^\dagger c_{x-\theta_n+2q\pi},\end{aligned} \tag{B2}$$

where now the sum over $q$ is only for $q$ such that $x - y + \theta_n - 2q\pi = 0$ is possible. One can show that this is only possible for $q = 0$ and $q = 1$. Thus the real-space operators are

$$T_n = \int_{-\pi}^{\pi} \mathrm{d}x\, c_x^\dagger [c_{x-\theta_n} + c_{x+2\pi-\theta_n}]. \tag{B3}$$

For a one-body state, each term in this expression can be interpreted as rotating the fermion's position from one place on the circle to another; for a many-body state, every fermion is moved in this way.

## APPENDIX C: REAL-SPACE COMMUTATOR

Consider now the real-space commutator, i.e., the commutator of the real space expressions for $T_n$ and $H_{\text{int}}$. In a similar manner, consider

$$[c_x^\dagger c_{x+\theta_n}, c_{q+p}^\dagger c_q c_k c_{k-p}^\dagger], \tag{C1}$$

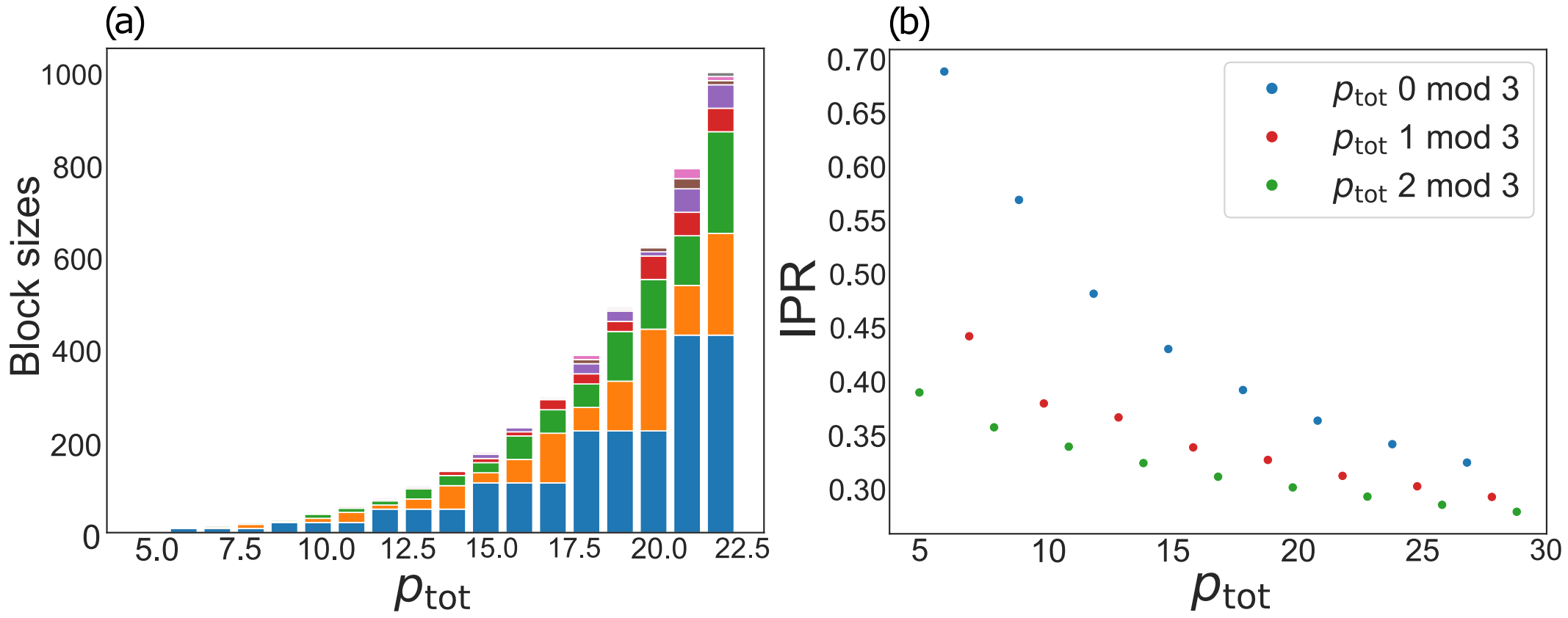

FIG. 10. Same as Fig. 5 but for symmetry $T_3$. (a) Stacked-bar histograms of block sizes. At a given $p_{\rm tot}$ there are more blocks for symmetry $T_3$ than $T_2$. (b) IPR against $p_{\rm tot}$. The IPR still tends to 0 for all patterns in the family. We color code the $p_{\rm tot}$ off different moduli with respect to $n = 3$.

which by using real-space anticommutation relations yields

$$\delta(x+\theta_n-y)c_x^\dagger c_y c_z^\dagger c_z - \delta(x+\theta_n-z)c_x^\dagger c_y^\dagger c_y c_z + \delta(x-y)c_y^\dagger c_z^\dagger c_z c_{x+\theta_n} - \delta(x-z)c_y^\dagger c_y c_z^\dagger c_{x+\theta_n}. \tag{C2}$$

Computing $[T_n, c_y^\dagger c_y c_z^\dagger c_z]$ amounts to integrating this over $x$. Rearranging two of the terms so all indices are of the form $r \pm \theta_n$ within the first two operators yields

$$c_{y-\theta_n}^\dagger c_y c_z^\dagger c_z + c_{z-\theta_n}^\dagger c_z c_y^\dagger c_y - c_y^\dagger c_{y+\theta_n} c_z^\dagger c_z - c_z^\dagger c_{z+\theta_n} c_y^\dagger c_y + \delta(z-y-\theta_n)c_y^\dagger c_z + \delta(y-z-\theta_n)c_z^\dagger c_y - \delta(y-z)c_{z+\theta_n}^\dagger c_y - \delta(y-z)c_y^\dagger c_{z+\theta_n}. \tag{C3}$$

Now multiplying by $\frac{1}{2}V(y-z)$ and integrating over two copies of $S^1$ in $y, z$ yields the full commutator $[T_n, H_{\rm int}]$ as

$$\frac{1}{2}\int {\rm d}y{\rm d}z\, V(y-z)[c_{y-\theta_n}^\dagger c_y c_z^\dagger c_z - c_y^\dagger c_{y+\theta_n} c_z^\dagger c_z] + \frac{1}{2}[V(\theta_n) - V(0)]\int {\rm d}y\,[c_y^\dagger c_{y+\theta_n} + c_{y-\theta_n}^\dagger c_y].$$

Now changing variables to $y' = y + \theta_n$ for the second term of the first integral and the second term of the second integral, using periodic boundary conditions and relabelling $y' \to y$ gives

$$\frac{1}{2}\int {\rm d}y{\rm d}z\,[V(y-z) - V(y-z-\theta_n)]c_{y-\theta_n}^\dagger c_y c_z^\dagger c_z + \frac{1}{2}[V(\theta_n) - V(0)]\int {\rm d}y\, c_{y-\theta_n}^\dagger c_y. \tag{C4}$$

For this commutator to vanish, we seem to need $V(x+\theta_n) = V(x)\ \forall x \in S^1$. In particular it suggests $V(0) = V(\theta_n)$, which when Fourier transforming implies

$$\sum_{\text{p not multiple of n}} V(p)(e^{i\theta_n p} - 1) = 0, \tag{C5}$$

which, e.g., for $n = 3$ gives

$$V(1)(e^{i2\pi/3} - 1) + V(2)(e^{i4\pi/3} - 1) + V(4)(e^{i2\pi/3} - 1) + V(5)(e^{i4\pi/3} - 1) + \cdots = 0, \tag{C6}$$

which is only true if both $V(1) + V(4) + \cdots = 0$ and $V(2) + V(5) + \cdots = 0$. If we use pattern $P_3$, we then minimally require $V(1) = V(2) = 0$ (see Fig. 10 for the corresponding block sizes). For general $n$, we require $V(1) = V(2) = \cdots = V(n-1) = 0$. This agrees with the momentum-space commutator, with the extra requirement that every term in $P_n$ be set to 0. We believe this results from the absence of normal-ordering of the Hamiltonian in the real-space derivation.

## APPENDIX D: NUMERICAL METHODS

We generate basis states $|\phi_i\rangle \in \mathcal{B}$ of a given momentum $p_{\rm tot}$ by using the integer partitions of $p_{\rm tot}$. For more details, see [31]. Any many-body state of total momentum $p_{\rm tot}$ can be obtained from the ground state $|\Omega\rangle$ by creating excitations at momenta in the interval $[-p_{\rm tot}+1, p_{\rm tot}]$. Thus we can represent the basis states $|\phi_i\rangle$ by using a binary string of length $2p_{\rm tot}$. We call this mapping $b : \mathcal{B} \mapsto \{\text{binary strings of length } 2p_{\rm tot}\}$.

In this basis, we must then compute all terms of the form $\langle\phi_i|H_{\rm int}|\phi_j\rangle$. This corresponds to using binary operations AND and XOR between two basis states $b(|\phi_i\rangle)$, $b(|\phi_j\rangle)$ to compute the changes in positions of the fermions. Our interaction Hamiltonian is a two-body interaction, so states differing at more than four momentum sites cannot scatter. The resulting scattering terms for $i \neq j$ have the form $V(p) - V(q)$ because there are always two ways of moving fermions from their initial to final sites, with a minus sign between them guaranteed by the exchange statistics for fermions. Figure 11 shows an example between two basis states of $p_{\rm tot} = 3$, as described in Sec. II. The resulting scattering term between $|\phi_3\rangle$ and $|\phi_2\rangle$ is $V(3) - V(1)$. For $i = j$, we have all scatterings that move pairs without changing the overall state.

To the resulting matrix, we add the diagonal kinetic terms to obtain the full matrix $\langle\phi_i|H|\phi_j\rangle$. We use sparse matrices to avoid storing unnecessary zeros between states that cannot scatter. A BFS algorithm is sufficient to determine whether a given matrix can be permuted into block-diagonal form.

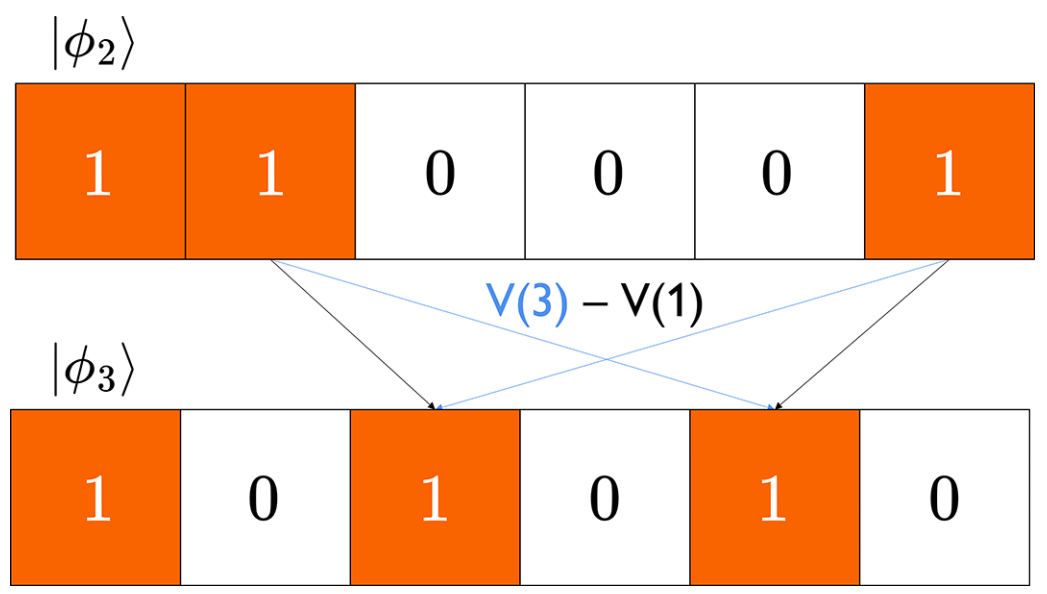

FIG. 11. Example of a scatter computation between two states. States are represented by binary strings indicating which momenta are occupied (1s) and which are not (0s). Scattering between two states uses the XOR operation between these two binaries, and finds the two possible permutations to get the scattering term, here $V(3) - V(1)$.

## APPENDIX E: BLOCK SIZE SEQUENCE: $T_2$

In Sec. III B we state that the $T_2$ block size sequence follows $a_2(j)$. More specifically, let $b_\lambda(p_{\rm tot})$ be the number of states in a block $b$ labeled by $T_2$ eigenvalue $\lambda$ in momentum subsector $p_{\rm tot}$. If the eigenvalue first arises in subsector $p_{\rm tot} = t$, then we have

$$b_\lambda(t + 2j) = a_2(j) \text{ for } j \in \mathbb{Z}_0^+, \tag{E1}$$

which also implies $b_\lambda(t) = 1$; a new block always starts with a single state. We also recall that $t$ is a triangular number.

The sequence $a_2(j)$ (A000712 in the OEIS) is known to be equal to the number of partitions of $2j$ in which the odd numbers appear as many times in odd as even positions. For example, setting the convention that all partitions are in descending order, we can take the partitions

$$[4^{(1)}, 2^{(2)}] \text{ and } [3^{(1)}, 1^{(2)}, 1^{(3)}, 1^{(4)}], \tag{E2}$$

of 6, where each number has been labeled by its position in the partition. Both partitions satisfy the requirement, as odd numbers appear in as many odd as even positions. Calling this sequence $p_{\rm oe}(2j)$ and the set of all such partitions $S_{\rm oe}(2j)$, we have that

$$a_2(j) = p_{\rm oe}(2j),$$

and we hope to show that

$$S_{\rm oe}(2j) \overset{\text{bij.}}{\Longleftrightarrow} \text{states at } p_{\rm tot} + 2j \text{ with same}$$
$$n_{\rm e} - n_{\rm o} \text{ value.} \tag{E3}$$

Consider the block with $n_{\rm e} - n_{\rm o} = 0$; the first state to satisfy this is $|\Omega\rangle$ at $p_{\rm tot} = 0$. Suppose we are interested in all the states at $p_{\rm tot} = 6$, which also satisfy $n_{\rm e} - n_{\rm o} = 0$—we want to take partitions in $S_{\rm oe}(6)$ and use them to map from $|\Omega\rangle$ to these states. We do this by moving electrons according to the numbers in the partition, starting with the rightmost; for example [4,2] corresponds to moving the rightmost ($p = 0$) electron in $|\Omega\rangle$ four spaces to the right, then the next rightmost two to the right.

This guarantees that partitions in $S_{\rm oe}$ will not upset the $n_{\rm e} - n_{\rm o}$ value. Moving an electron an even distance will clearly never change $n_{\rm e} - n_{\rm o}$. The fact that odd numbers appear in as many odd as even positions means that every odd site → even site movement is canceled out by an even → odd movement, as in Fig. 12. In fact, of the 11 partitions of six, only the "triangular partition" [3,2,1] does not satisfy the requirement to be in $S_{\rm oe}$. This partition maps $|\Omega\rangle$ to $c_3^\dagger c_1^\dagger c_0 c_{-2}|\Omega\rangle$, which marks the beginning of the $n_{\rm e} - n_{\rm o} = -4$ block; the growth of this block is then described by the same sequence.

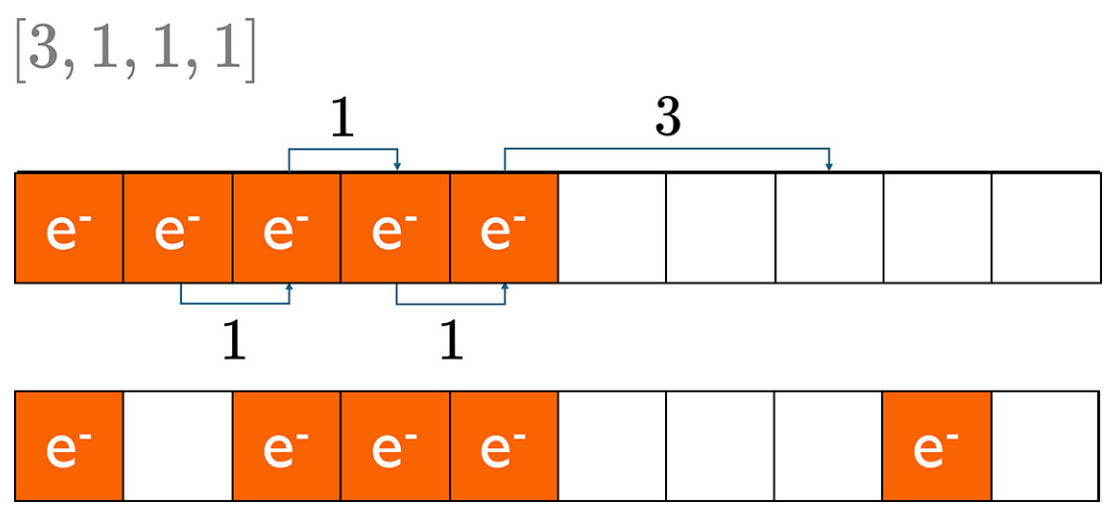

FIG. 12. Representation of how states with the same $n_{\rm e} - n_{\rm o}$ numbers can be generated from one another via partitions in the set $S_{\rm oe}$. Here, $c_3^\dagger c_{-3}|\Omega\rangle$ (second row) is created from $|\Omega\rangle$ (first row) via the partition [3, 1, 1, 1] in $S_{\rm oe}(6)$. Both have $n_{\rm e} - n_{\rm o} = 0$ because the partition is in $S_{\rm oe}$.

This is therefore an alternate method of generating states, which immediately classifies them by their $T_2$ eigenvalues. Counting the relative frequency of odd numbers in odd vs even positions *in the partition* is completely analogous to counting odd vs even site occupancies *in the state*. Since the number of states at $p_{\rm tot}$ is the number of partitions of $p_{\rm tot}$, the mapping described must be a bijection.

This convention works because of the form of $|\Omega\rangle$; we start from the rightmost electron so we first move an even ($p = 0$) electron, then an odd and so on. It must be modified for the triangular states, which begin the other blocks; consecutive odd numbers in the partitions in $S_{\rm oe}(2j)$ must be applied to the leftmost holes, moving them left, instead of the rightmost electrons.

Finally we note that $p_{\rm oe}(2j + 1) \equiv 0$. This is because in any partition of an odd number $2j + 1$, if there are $n_{\rm e}$ and $n_{\rm o}$ even and odd numbers respectively, we must have that $n_o$ is odd. But then there cannot be as many odd numbers in even as odd positions—this requires $n_{\rm o}$ even. This means that there is no way of generating a state in the $p_{\rm tot} + 2j + 1$ subspace from one in the $p_{\rm tot}$ subspace without changing the $T_2$ eigenvalue. This is an alternate way of justifying that odd and even $p_{\rm tot}$ values should be considered separately.

We emphasise again that the numerical results imply the above generalises to $T_n$ for $n$ prime, but for now this remains a conjecture. For $n > 2$ we also do not know how to predict the values of $p_{\rm tot}$ at which new blocks arise; for $n = 3$ these values yield an interesting sequence, but for higher $n$ we quickly find that new blocks ($T_n$ eigenvalues) arise in almost every $p_{\rm tot}$ subsector.

## APPENDIX F: POTENTIAL CONFIGURATIONS

To realize various potentials $V(x)$, we work with Fourier Transform components

$$V(p) = \frac{1}{L}\int_{-L/2}^{L/2} dx\, e^{ipx} V(x), \tag{F1}$$

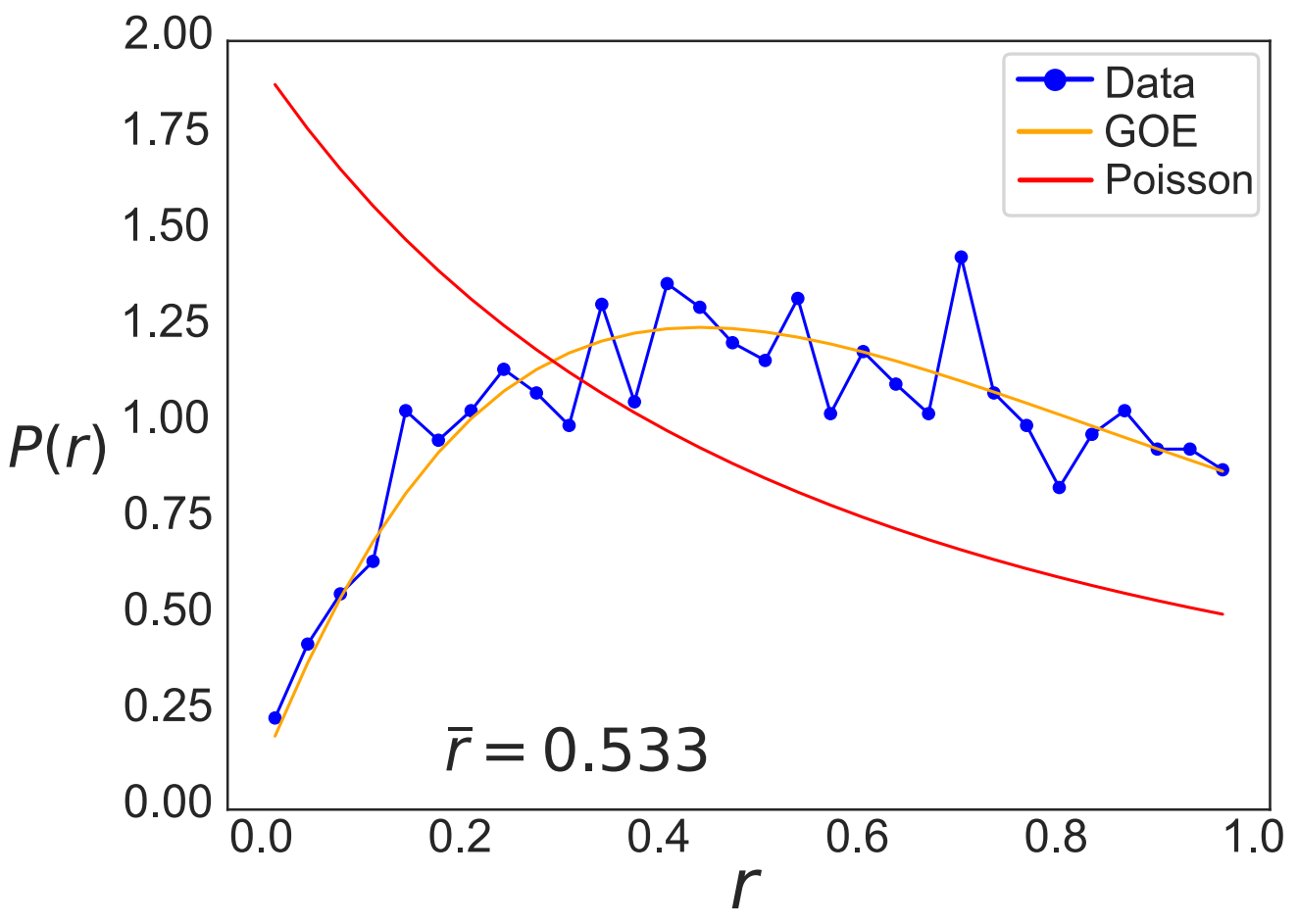

FIG. 13. Adjacency gap ratio distribution, see Sec. III C. Numerical parameters are as in Fig. 1. Drawn in red is the Poisson distribution; in orange is the GOE distribution. Blue points represent level statistics within the biggest Hamiltonian block. The data closely follow the GOE distribution, with $\bar{r} \approx 0.533$, indicating thermalization.

where again $p \in \mathbb{Z}$ owing to periodic boundary conditions. Following Ref. [21], we investigate various potential configurations by setting equal a subset $S$ of all $V(p)$. As described in Sec. II, this approach allows the determination of minimal tuning patterns that yield exact states. The remaining components are drawn randomly from a uniform interval, to make the matrix as "thermal" as possible, i.e., making it follow the ETH as closely as possible. Throughout this paper, we choose all $V(p)$ to be repulsive, i.e., $V(p) > 0$. Attractive $V(p) < 0$ gives boson condensation.

Compared to Ref. [21], we make one slight modification. While that reference used a quadratic energy dispersion relation $\epsilon(p) = vp + ap^2$, we add a cubic term $bp^3$ to ensure that the condition $\text{sgn}(\epsilon(p)) = \text{sgn}(p)$ is true for all $p$. In [21], the reasoning behind the quadratic term was because it was the first move away from the integrable bosonic limit $\epsilon(p) = vp$. Moreover, with a small enough coefficient $a$, the odd nature of $\epsilon(p)$ would only be compromised for very negative $p$, thus only affecting high-energy states. Indeed adding the (small) cubic term and comparing energy spectra to those in [21], we find that only the high-energy edge is affected.

## APPENDIX G: LEVEL STATISTICS

Here, we show the level statistics for the pattern $P_2$ and $p_{\text{tot}} = 30$ (Fig. 13), both for the full Hamiltonian and for the first (biggest) block.